\documentstyle[aps,prc,preprint,tighten,epsf]{revtex}
\begin{document}

\draft

\title{Correlations and the relativistic structure of the nucleon
self-energy}

\author{H.\ M\"{u}ther, S.\ Ulrych and H.\ Toki}
\address{Institut f\"ur Theoretische Physik, Universit\"at T\"ubingen,
         D-72076 T\"ubingen, Germany}

\maketitle

\begin{abstract}
A key point of Dirac Brueckner Hartree Fock calculations for nuclear matter is
to decompose the self energy of the nucleons into Lorentz scalar and vector
components. A new method is introduced for this decomposition. It is based on
the dependence of the single-particle energy on the small component in the
Dirac spinors used to calculate the matrix elements of the underlying NN
interaction. The resulting Dirac components of the self-energy depend on the
momentum of the nucleons. At densities around and below the nuclear matter
saturation density this momentum dependence is dominated by the non-locality of
the Brueckner G matrix. At higher densities these correlation effects are
suppressed and the momentum dependence due to the Fock exchange terms is
getting more  important. Differences between symmetric nuclear matter and
neutron matter are discussed. Various versions of the Bonn potential are
considered.
\end{abstract}


\section{Introduction}
The success of the phenomenological Walecka model\cite{wal1,serot} has motivated 
various attempts to account for relativistic effects in microscopic nuclear
structure calculations which are based on realistic nucleon-nucleon
interactions\cite{anast,rupr,brock,bm2,malf1,weigel}. These attempts, which are
often called relativistic Brueckner-Hartree-Fock or Dirac-Brueckner-Hartree-Fock
(DBHF) calculations, lead to a significant progress in describing the saturation
properties of symmetric nuclear matter. Conventional non-relativistic
Brueckner-Hartree-Fock calculations or applications of other techniques to
account for correlation effects typically yield a prediction for the saturation
point which occurs at a too large density and/or with too little binding energy
as compared to the empirical values. This desease of conventional many-body
calculations can be cured by introducing three-nucleon forces\cite{bwi}. DBHF
calculations, however, reproduce the empirical saturation properties of nuclear
matter without the need to introduce a three-nucleon force\cite{brock}.

This success of the DBHF approach does not originate only from the use of
relativistic kinematics. The key-point is the decomposition of the
single-particle potential or the nucleon self-energy into two large components,
a scalar part $\Sigma_s$ and a time-like component of the vector part 
$\Sigma_0$. Each of these components is quite large with absolute values of a
few hundred MeV. This means they are of a size comparable to the nucleon rest
mass. Employing a self-energy with this relativistic structure in a Dirac
equation for the nucleons in nuclear matter yields nucleons, which are
weakly bound, which is in agreement with the empirical
fact that the binding energy of a nucleon in nuclei is very small as compared to
the rest mass of the nucleon. This is because the  two
large components compensate each other to a large extent in calculating binding
energies. The mean-field single-particle energies derived from relativistic and
non-relativistic calculations are rather similar.

The large scalar component $\Sigma_s$ in the relativistic self-energy, however,
leads to Dirac spinors which are quite different from the Dirac spinors of free
nucleons. In solving the Dirac equation $\Sigma_s$ can be combined with the 
mass of
the nucleon leading to an effective Dirac mass $m_D^*$, which is significantly
smaller than the bare nucleon mass. This means that the Dirac spinors 
for the nucleons in the
medium show a small component which is enhanced as compared to the free Dirac 
spinor. The matrix elements of the nucleon-nucleon (NN) interaction calculated
within the meson exchange model depend on the structure of the Dirac spinor.
This means that the NN interaction between two nucleons within the medium of
nuclear matter is different from the interaction of two corresponding nucleons
in vacuum. This medium dependence of the NN interaction, which can be
characterized by the effective Dirac mass $m_D^*$, is the essential ingredient
of DBHF calculations which has lead to the improvement as compared to
non-relativistic calculations. 

The analysis of the nucleon self-energy in terms of its relativistic components 
is a very important step in such relativistic calculations. Within the framework
of DBHF this analysis is non-trivial. The DBHF approach accounts for the
effects of correlations by solving a three-dimensional Bethe-Goldstone equation.
This leads to matrix elements of the so-called G-matrix just between the Dirac
states of positive energy. From this G-matrix one can 
evaluate the single-particle energy for a nucleon with momentum $k$. The problem
is how to derive from the G-matrix or the momentum dependence of the
single-particle energy, which is just an information about the positive energy
Dirac spinors, the relativistic structure of the nucleon self-energy.

One possibility, which has been used e.g.~by Brockmann and Machleidt\cite{brock}
and Engvik et al.\cite{engvi},
is to extract the Dirac structure of the self-energy from the momentum
dependence of the single-particle energy. One assumes that the structure of the
self-energy is identical to the one obtained within the mean-field or
Dirac-Hartree approach for nuclear matter. This means that the space-like
vector component $\Sigma_v$ vanishes and the time-like vector component
$\Sigma_0$ as well as the scalar component $\Sigma_s$ are constant,
i.e.~independent of the nucleon momentum $k$. These two constants are then
adjusted to reproduce momentum dependence of the calculated DBHF single-particle
energy spectrum $\epsilon (k)$. Attempts have been made to extend this scheme 
and to allow for momentum dependent components $\Sigma_s$ and
$\Sigma_0$\cite{lee}. It is evident, however, that it is impossible to derive
two functions $\Sigma (k)$ from one observable in an unambigous way. 

There are various sources which define the momentum dependence of the 
single-particle energy or the non-locality of the single-particle potential. The
relativistic effects contained in a Dirac-Hartree approximation represent one of
the possible sources for this non-locality, which is often parametrized in terms
of an effective mass. Other sources for the non-locality are the Fock exchange
contributions to the mean field and the non-locality of the effective
NN interaction, the G matrix. A careful discussion of these various
contributions has been given by Jaminon and Mahaux\cite{jami}. It is the aim of
the present work to disentangle, these various contributions within the 
framework of DBHF. 

Another scheme to extract the Dirac structure of the nucleon self-energy, 
which has been used by various groups\cite{horow,sehn,dejong},
is based on the attempt to analyze the matrix elements of the G-matrix in 
terms of a relativistic operator for a two-body interaction. Five amplitudes
multiplying the different invariant operators are adjusted to reproduce the
G-matrixelements. The Dirac structure of the self-energy can then be derived
from these amplitudes. Also this procedure, however, is model-dependent. The
results depend on the chosen set of invariants. Furthermore the DBHF calculation
only yields matrixelements between antisymmetrized two-nucleon states. The
decomposition of these matrixelements into direct and exchange contributions,
which is required for this analysis, cannot be obtained in a unique way.

To avoid such ambiguities one has to calculate the scattering amplitude or G
matrix in the complete Dirac space and not only for the positive energy spinors.
Such an approach has been used by Huber et al.~\cite{weigel} using the so-called
$\Lambda$-approximation. In the investigation presented here, we would like to
perform a step into this direction by evaluating the G matrix and the resulting
single-particle energy for positive energy Dirac spinors with different 
decompositions into small and large components. The enhancement of the small
component is characterized by the
Dirac mass $m_D^*$. From the dependence of the two-body interaction and the
single-particle energy on this parameter $m_D^*$, i.e.~on the ratio of small to
large component we can derive the Dirac structure of the underlying nucleon
self-energy. 

This technique will be applied to perform DBHF calculations for nuclear matter
and neutron matter employing various versions of the Bonn potential\cite{rupr}.
These results are compared to DBHF calculations, in which the Dirac structure of
the self-energy has been deduced from the momentum dependence of the
single-particle potential\cite{brock}. We will pay special attention to the
influence of correlation effects on the momentum dependence of the self-energy.
The results of this analysis will also be
used to derive an effective NN interaction, depending on density and asymmetry
of the nuclear system, to be used in Dirac-Hartree-Fock calculations of finite
nuclei\cite{fritz,toki1,toki2,boer1,boer2}. 
After this introduction we will review some of the basic ingredients of
the DBHF formalism in section 2 and define the technique to determine the Dirac
structure of the self-energy. Numerical results are presented in section 3 and
the main conclusions are summarized in section 4.

\section{Dirac Brueckner Hartree Fock Theory}
In isotropic nuclear matter the self-energy of a nucleon with momentum $k$
can be decomposed in the following components of the spinor representation
\begin{equation}
\Sigma (k) = \tilde{\Sigma_s} (k) - \gamma^0 \tilde{\Sigma_0} (k) + \gamma \cdot
{\bf k} \Sigma_v (k) \, .
\label{eq:selfen1}
\end{equation}
All other components vanish because of isotropy of the system. Inserting this
self-energy into the Dirac equation for a nucleon in the nuclear
medium, we obtain
\begin{equation}
\left[ \left( 1+\Sigma_v (k)\right) \gamma \cdot {\bf k} + \left( m +
\tilde{\Sigma_s} (k) \right) - \tilde{\Sigma_0} (k)  \gamma^0 \right]
u (k) = \tilde\epsilon (k) \gamma^0 u (k)
\label{eq:dirac1}
\end{equation}
Now it is convenient to eliminate the space-like vector component $\Sigma_v$ 
and rewrite this Dirac equation
into a form which only contains a scalar and a time-like vector component
\begin{equation}
\left[  \gamma \cdot {\bf k} + \left( m +
{\Sigma_s} (k) \right) - {\Sigma_0} (k)  \gamma^0 \right]
u (k) = \tilde\epsilon (k) \gamma^0 u (k)
\label{eq:dirac2}
\end{equation}
where
\begin{eqnarray}
\Sigma_s & = & \frac{\tilde{\Sigma_s} - m \Sigma_v}{1+\Sigma_v}
\nonumber \\
\Sigma_0 & = & \frac{\tilde{\Sigma_0} - \epsilon \Sigma_v}{1+\Sigma_v}
\label{eq:efself}
\end{eqnarray}
Comparing the Dirac equation as displayed in (\ref{eq:dirac2}) with the Dirac
equation for a free nucleon, it is obvious that the solution of this Dirac
equation for a nucleon in the medium is identical to the one for a free nucleon
if we replace the bare mass $m$ by the effective Dirac mass
\begin{equation}
m_D^* = m + {\Sigma_s} (k)
\label{eq:mdirac}
\end{equation}
This implies that the Dirac spinors which are obtained as solutions of this
Dirac equation for states with positive energy take the form
\begin{equation}
u_\alpha (k) = \sqrt{\frac{E_k^*+m_D^*}{2E_k^*}}\left(\begin{array}{c}
 1 \\ \frac{\vec{\sigma}\cdot \vec
k}{E^*_k+m^*_{D}}\end{array}\right)\chi_\alpha
\label{eq:uspin}
\end{equation}
where $\alpha$ refers to the spin of the nucleon and $\chi_\alpha$ represents
the corresponding Pauli spinor. Note that this spinor is normalized in such a
way that $u^\dagger u = \bar{u} \gamma^0 u = 1$. The constant $E_k^*$ stands for 
$$
E_k^* = \sqrt{k^2 + {m_D^*}^2}
$$ 
and can also be used to define the time-like component or single-particle energy
of the four-momentum $k$
$$ k^0 = \epsilon (k) = E_k^* - \Sigma_0$$
This single-particle energy can also be calculated as an ``expectation value''
of the Dirac operator
\begin{eqnarray}
\epsilon (k) & =  \bar u \left[ \gamma \cdot {\bf k} + m \right] u & +
\bar u \Sigma u \nonumber \\
& = T (k) & + U(k) \label{eq:tplusu}
\end{eqnarray}
with a contribution $T(k)$ originating from the free part of the Dirac operator
\begin{equation}
T(k) = \frac{m m_D^* + k^2}{E_k^*} \label{eq:tkin}
\end{equation}
and the expectation value for the self-energy, the single-particle potential
\begin{equation}
U(k) = \frac{m^*_D}{E_k^*}\Sigma_s - \Sigma_0\, .
\label{eq:uofk}
\end{equation}
Assuming that the NN interaction can be described in terms of a meson-exchange
or One-Boson-Exchange (OBE) model, the self-energy can easily 
be calculated in the Dirac-Hartree-Fock approximation. As an example we consider
the exchange of $\sigma$ and $\omega$ and show the
expression for the scalar part of the self-energy, which can be written as
\begin{eqnarray}
\tilde\Sigma_s(k)  & = & - \left(\frac{g_\sigma}{M_\sigma}\right)^2 \rho_s \quad +
\quad \frac{1}{(4\pi)^2}\frac{1}{k}\int_0^{k_F} p\, dp\frac{m_D^*}{E_p^*} \times
\nonumber \\
&&\left[g_\sigma^2 \Theta_\sigma(k,p) - 4 g_\omega^2 \Theta_{\omega} (k,p) 
\right] \label{eq:sigs}
\end{eqnarray}
$g_\sigma$ and $g_\omega$ represent the meson-nucleon coupling
constants for these  mesons and $M_\alpha$ represent the corresponding
masses. The first term on the right hand side of (\ref{eq:sigs}) originates from
the Hartree contribution and contains the scalar density
$$
\rho_s = \frac{2}{\pi^2}\int_0^{k_F} p^2\, dp\, \frac{m_D^*}{E_p^*}
$$ 
with $k_F$ the Fermi momentum. The functions $\Theta_\alpha$ occuring in the
Fock (exchange) contributions are defined by
$$
\Theta_\alpha(k,p) = \ln \left(\frac {k^2+p^2+M_\alpha^2 - (\epsilon (k)-
\epsilon(p))^2+2pq}{k^2+p^2+M_\alpha^2 - (\epsilon (k)-\epsilon(p))^2 - 2pq}
\right)
$$
Similar expression are obtained for the other components of the self-energy
$\tilde\Sigma_0$ and $\Sigma_v$. They are listed e.g.~in reference
\cite{fritz}. 

Using the Hartree approximation, $\tilde\Sigma_s$ is a constant, independent of
the nucleon momentum $k$ and originates completely from the $\sigma$ exchange
part of the NN interaction. Within this approximation $\Sigma_v$ is identically
zero and also the space-like vector component $\tilde\Sigma_0$ is constant.
Within the Dirac-Hartree-Fock approximation, the momentum dependence of the
various components in the nucleon self-energy originates from the various Fock
exchange terms like those in the second term of (\ref{eq:sigs}). 

The expressions for the components of the self-energy depend on the effective
Dirac mass $m_D^*$. This dependence is due to the fact that  
the meson-exchange matrix elements, which are needed to determine the nucleon
self-energy, are calculated using the Dirac spinors, which arise from the
solution of the Dirac equation (\ref{eq:dirac2}). The matrix elements of the NN
interaction in the medium, calculated for the dressed nucleon spinors, are
different from the corresponding ones in the vacuum. These differences are
characterized by the effective Dirac mass $m_D^*$. It is obvious that the   
solution of the Dirac equation and the calculation of the nucleon self-energy
requires a self-consistent procedure.

The Dirac-Hartree-Fock approximation, however, is not a useful approximation to
describe nuclear properties if one considers realistic meson-exchange 
potentials, which are adjusted to fit NN scattering data. It would yield unbound
nuclei (see also below). The origin of this disease is the existence of the
strong short-range and tensor components in such a realistic NN interaction,
which must be accounted for by considering the effects of short-range NN
correlations in the nuclear wave function. This can be done using the tools of
the Brueckner Hartree Fock approximation.

As a first step of such a Dirac-Brueckner-Hartree-Fock approximation, one
evaluates the matrix elements of the meson-exchange potential 
\begin{equation}
<kp\vert V(m_D^*) \vert k'p'>  = \bar{u}(k) \bar{u}(p) {\cal V} u(k') u(p')
\label{eq:vnn}
\end{equation}
with ${\cal V}$ representing the relativistic operator for the
OBE approximation of
the NN interaction. The medium dependence of these matrix elements originating
from the medium dependence of the Dirac spinors $u$ (see eq.(\ref{eq:uspin}))
is emphasized by the formal parameter $m_D^*$ on the left hand side of
(\ref{eq:vnn}). These matrix elements may now be used, like matrix elements of a
conventional non-relativistic NN interaction, to solve the Bethe-Goldstone
equation
\begin{eqnarray}
<kp \vert G(m_D^*,\Omega ) \vert kp> & = &<kp\vert V(m_D^*) \vert kp> +
\int d^3p' \int d^3k' <kp\vert V(m_D^*) \vert k'p'> \times \nonumber \\&&
\qquad
\frac{Q(k',p')}{\Omega -
\epsilon_{k'}-\epsilon_{p'}} <k'p' \vert G(m_D^*,\Omega ) \vert kp>
\label{eq:betheg}
\end{eqnarray}
with the starting energy parameter $\Omega$, the Pauli operator $Q$ and the
single-particle energies for the intermediate states $\epsilon_{k'}$,
$\epsilon_{p'}$. If one ignores the medium effects expressed in terms of the
Pauli operator, the dressed nucleon spinors ($u$, $m_D^*$) and energies and
replaces these by the corresponding quantities of the free nucleon, this
Bethe-Goldstone equation becomes  the Thompson equation, a specific
three-dimensional reduction of the Bethe-Salpeter equation for NN scattering.
Since the three versions of the Bonn potential defined in table A.2 of
\cite{rupr} have been fitted to the NN scattering data using this Thompson
equation, it is consistent to use the Bethe-Goldstone 
equation in the form just outlined
in order to determine the effective NN interaction of two nucleons in the
medium, which accounts for NN correlations. Note, however, that this procedure
leads to matrix elements between nucleon states and does not keep track of 
any Dirac structure.

The effective interaction $G$ can then be used to determine the single-particle
potential of the nucleons in nuclear matter by
\begin{equation}
U(k) = \int_0^{k_F} d^3p \,  <kp \vert G(m_D^*,\Omega =
\epsilon_{k}+\epsilon_p ) \vert kp>  
\label{eq:ubhf}
\end{equation}
The problem is how to determine the Dirac structure of the self-energy from this
function $U(k)$. One possibility is to assume that the self-energy components
preserve the properties of the Dirac-Hartree approach, i.e.~$\Sigma_v$
vanishes and $\Sigma_s$ and $\Sigma_0$ are constant, independent on the momentum
of the nucleon. In this case one can identify the potential energy of
(\ref{eq:ubhf}) with (\ref{eq:uofk}) and determine the two constants $\Sigma_s$
and $\Sigma_0$ such that the Dirac-Hartree expression (\ref{eq:uofk}) fits the
potential energy derived in the DBHF approach of (\ref{eq:ubhf}). With this
Dirac-Hartree assumption it is also consistent to parametrize the starting
energy $\Omega$ and the single-particle energies $\epsilon_k$ occuring in the
propagator of the Bethe-Goldstone eq.(\ref{eq:betheg}) in the same way
\begin{equation}
\epsilon_k = \sqrt{k^2+{m_{BG}^*}^2} -\Sigma_0
\label{eq:mbg}
\end{equation}
using the same effective mass parameter
\begin{equation}
m_{BG}^* = m_D^* = m + \Sigma_s
\label{eq:brok}
\end{equation}
An iterative procedure to solve these DBHF equations in a self-consistent way
could proceed as follows: Assume a value for $m_D^*$ - calculate the matrix
elements of the NN interaction for the corresponding spinors following
(\ref{eq:vnn}) - solve the Bethe-Goldstone eq.(\ref{eq:betheg}) using the same
effective mass to determine the single-particle spectrum - calculate $U(k)$
according (\ref{eq:ubhf}) and determine a new value for $\Sigma_s$ which
redefines $m_D^*$. Repeat these steps until convergence is achieved.

This procedure, which we will refer to in the following as the Dirac-Hartree
(DH) self-consistency scheme has been used by various groups, 
see e.g.~\cite{brock,engvi}. There seems to be a general consensus that the 
underlying assumption that $\Sigma_s$ and $\Sigma_0$ are essentially constant 
is fulfilled. Inspecting the momentum dependence of these self-energies, derived
from the analysis of the G matrix elements, which we discussed already in the
introduction\cite{horow,sehn,dejong,boer1}, indicates that the variation of
$\Sigma_s$ and $\Sigma_0$ with the momentum $k$ is weak on the scale of the
absolute values of these quantities (typically below 10
percent)\cite{ulr1,elsen}. Such variations, however, are large on the scale of
the single-particle potential $U(k)$ to which these two components add with
opposite sign. Therefore it can be very dangerous to ignore even a small 
momentum dependence in analyzing $U(k)$. It has been shown that applying such 
an analysis to asymmetric nuclear systems, including neutron matter, yields
isovector terms with even an opposite sign compared to those which were 
derived from a more detailed analysis.

Therefore we propose a method which improves this DH self-consistency scheme.
It yields momentum dependent components in the nucleon self-energy
and allows a separation of the exchange, correlation and relativistic effects
leading to the momentum dependence of the single-particle potential $U(k)$.
This is achieved by evaluating the single-particle potential of (\ref{eq:ubhf})
for various kinds of Dirac spinors characterized by the effective Dirac mass
$m_D^*$. This means that we calculate the single-particle potential, keeping the
value $m_D^*$ fixed, fulfilling only the Brueckner self-consistency requirement.
In the limit $m_D^*=m$ this procedure would correspond to the conventional BHF
approach, which ignores any change of the Dirac spinors in the medium
completely. In this limit of the conventional BHF approach we have to satisfy
just the BHF self-consistency condition, i.e.~determine the effective mass
$m_{BG}$ which parametrizes the single-particle spectrum $\epsilon_k$ to be used
in the Bethe-Goldstone equation in a self-consistent way. 

Employing this procedure for various values of $m_D^*$ we obtain
a function $U(k, m_D^*)$. At each nucleon momentum $k$ one can
analyze this potential $U$ as a function of the Dirac mass $m_D^*$ to probe the 
sensitivity of the calculated BHF potential on the structure of the underlying
Dirac spinors. To do so, we define for each $k$ two effective coupling
constants, $\Gamma_{\sigma}(k)$ and $\Gamma_{\omega}(k)$ for an effective 
scalar and vector meson, respectively. These coupling constants are adjusted to
reproduce at each momentum $k$ the dependence of the single-particle potential
on the  Dirac mass $m_D^*$ by
\begin{eqnarray}
U(k, m_D^*) & = & \int_0^{k_F} d^3p \,  <kp \vert G(m_D^*,\Omega =
\epsilon_{k}+\epsilon_p ) \vert kp> \nonumber\\
& = & - \Gamma_{\sigma}^2(k)\frac{2}{\pi^2}\int_0^{k_F} p^2\, dp\,
\frac{m_D^*}{\sqrt{p^2+m_D^{*2}}}\, +\, \Gamma_{\omega}^2(k)\rho
\label{eq:mdfit}
\end{eqnarray}
It turns out that the dependence of the DBHF single-particle potential (first
line of eq.(\ref{eq:mdfit})) on the parameter $m_D^*$ is very well reproduced by
the two parameter fit displayed in the second line of eq.(\ref{eq:mdfit}) which 
corresponds to the Dirac Hartree expression with coupling constants adjusted at
each momentum $k$. With these effective coupling constants $\Gamma (k)$ we can
now determine the self-consistent value for the effective Dirac mass by
\begin{equation}
m_D^* (k) = m + \Sigma_s (k) = 
m - \Gamma_{\sigma}^2(k)\frac{2}{\pi^2}\int_0^{k_F} p^2\, dp\,
\frac{m_D^*}{\sqrt{p^2+m_D^{*2}}}\, .
\label{eq:mdsts}
\end{equation}
Note, that this procedure requires at each density $\rho$ the fulfillment of two
self-consistency conditions: one is the self-consistent definition of the
single-particle energies to be used in the Bethe-Goldstone equation. This
single-particle spectrum can be characterized by the effective mass parameter
$m_{BG}^*$. The other one is the self-consistent determination of the Dirac
structure of nucleon spinors, which is represented by the effective Dirac mass
$m_D^*$. These two effective masses need to be
identical only, if the components of the self-energy were indeed independent of
the nucleon momentum.

\section{Discussion of results}

In this section we will consider two examples of realistic OBE potentials, the
potentials $A$ and $C$ which are defined in table A.2 of \cite{rupr}. Both of
them were adjusted to fit NN scattering phase shifts with good accuracy using
the Thompson equation to solve the NN scattering problem in the vacuum.
Therefore they are suited with the DBHF treatment outlined in the preceeding
section. The main difference between these two interactions is the strength of
the tensor force. The d-state probability calculated for the Bonn $A$ potential
is 4.47 percent while the Bonn $C$ potential yields 5.53 percent.

\begin{figure}
\epsfysize=8.0cm
\begin{center}
\makebox[16.0cm][c]{\epsfbox{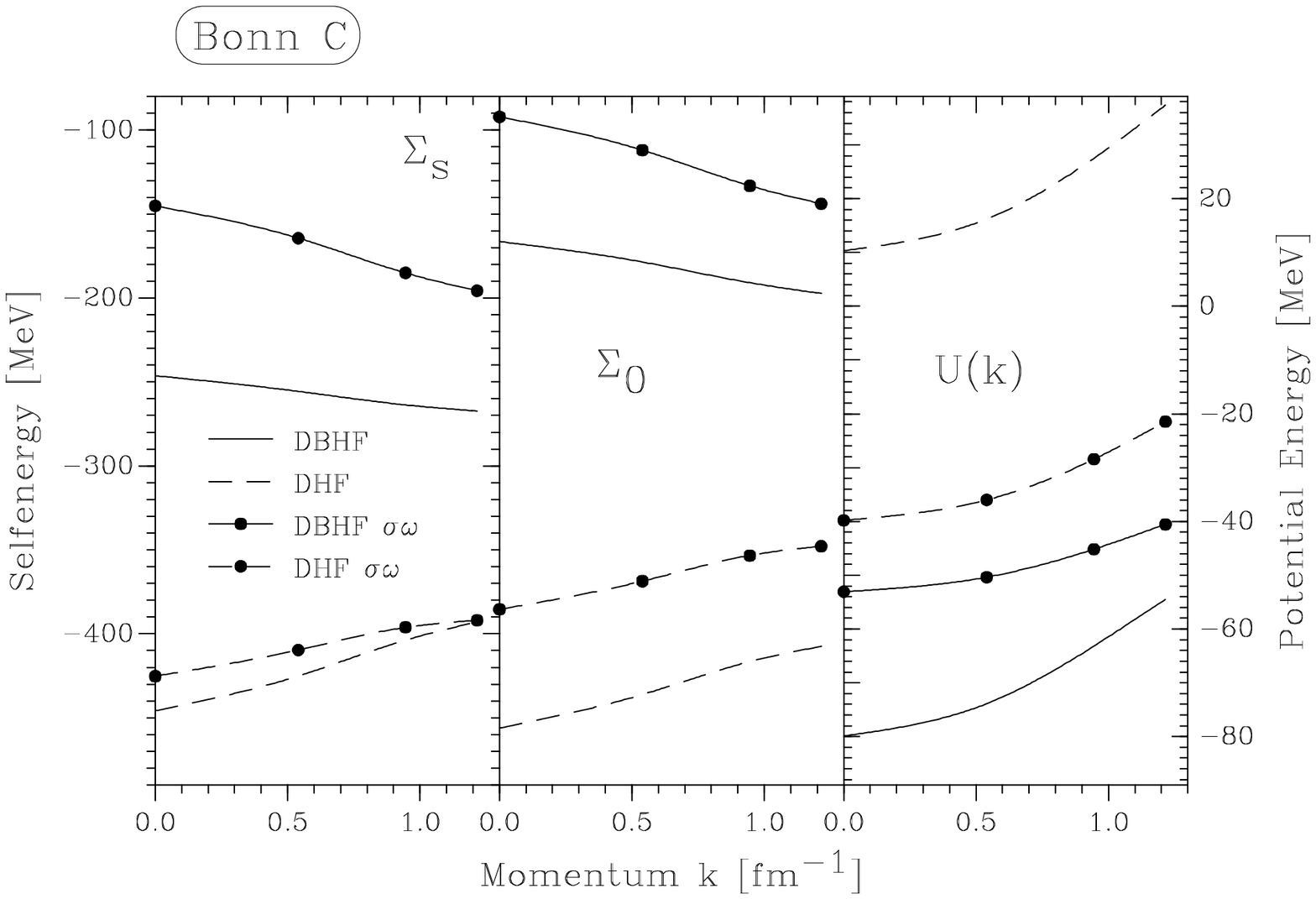}}
\end{center}
\caption{Momentum dependence of the scalar $\Sigma_s$ (left part) and vector
component $\Sigma_0$ (middle part) of the nucleon self-energy. Results of the
DBHF calculation are represented by solid lines, while the contribution of the
Born terms (denoted DHF) are shown by dashed lines. The Bonn $C$ potential has
been used in a calculation of nuclear matter with $k_F$ = 1.35 fm$^{-1}$ to 
derive these results. The curves marked with
little dots (DBHF $\sigma\omega$ and DHF $\sigma\omega$) were obtained in
calculations, in which only the $\sigma$ and $\omega$ exchange parts of Bonn $C$
were retained. The right part of the figure shows results for the
single-particle potential $U(k)$ calculated in the various approximations.}
\label{fig:km35c}
\end{figure}

Results of DBHF calculations for symmetric nuclear matter at the empirical
saturation density, which corresponds to a Fermi momentum $k_F$ of 1.35
fm$^{-1}$ using the potential $C$ are displayed in Fig.~\ref{fig:km35c}. The
left and the middle part of this figure show the scalar $\Sigma_s$ and the
vector part $\Sigma_0$ as a function of the nucleon momentum $k$. Note that we
consider the quantities which are renormalized according to
eq.~(\ref{eq:efself}), which includes effects of the space-like vector
component $\Sigma_v$. These two components show values ranging between -245 MeV
and -270 MeV for $\Sigma_s$ and  -165 MeV and -200 MeV for $\Sigma_0$
considering nucleon momenta between $k=0$ and the Fermi momentum. This means
that the variation of these quantities with the nucleon momentum is as large as
20 percent, which seems to be non-negligible. 

In order to explore this momentum dependence, we have calculated the
Dirac-Hartree-Fock (DHF) contribution to these components separetely.  This
means that we calculate and analyze in eq.(\ref{eq:mdfit}) the contribution 
to the single-particle
potential  $U(k, m_D^*)$, which is obtained replacing
the G matrix by the bare potential, i.e.~using the Born approximation. These
DHF contributions are presented in Fig.~\ref{fig:km35c} using dashed lines.
The  values of these DHF contributions are roughly twice as large as the total
terms. This demonstrates the importance of the short-range correlations, which
are taken into account using the G matrix.

The momentum dependence of these DHF terms is also non-negligible
(although the relative importance is decreasing as the absolute values are
larger), the variation of these terms with momentum, however, is just opposite
to the one obtained for the total DBHF. The decomposition of the DHF self-energy
shown in Fig.~\ref{fig:km35c} has been derived from the analysis of $U(k,
m_D^*)$ as outlined in the previous section. For the case of the DHF approach,
however, one can also determine the Dirac structure of the self-energy directly
using expressions like the example of (\ref{eq:sigs}). For the Bonn potentials
these expressions have to be extended to account for the formfactors and
all mesons included in the Bonn potential. The direct calculation and the
analysis of $U(k, m_D^*)$ yield identical results, a fact which has been used to
test our procedure of the analysis. This means that one can understand the
momentum dependence of the DHF contribution from analytic expressions like
(\ref{eq:sigs}). 

The increase of the DHF terms in the self-energy components
as a function of the momentum is mainly due to the interplay between the strong
exchange terms of the $\sigma$ and $\omega$ mesons. This can be seen from the
dashed lines marked with little dots which are obtained if only the
contributions of these two mesons are retained in evaluating $V$. Other mesons,
in particular the exchange of the rho meson, enhance the momentum dependence of
$\Sigma_s$ a bit and reduce the value of $\Sigma_0$.

As a next step we would like to discuss the effect of correlations within this
simplified $\sigma - \omega$ version of the Bonn OBE potential. The 
self-energy components calculated with a $G(\sigma ,\omega )$ which results
from the solution of the Bethe-Goldstone eq.(\ref{eq:betheg}) using  a NN
interaction $V$ which just contains the $\sigma$ and $\omega$ exchange from the
Bonn $C$ potential are presented by the solid lines marked with dots in
Fig.~\ref{fig:km35c}. Note that these results are not results of a
self-consistent DBHF calculation, the parametrization of the single-particle
energies ($m_{BG}^*$) and the Dirac spinors is identical to the one derived from
the DBHF for the complete NN interaction. 

The correlations reduce the absolute values of the self-energy in a very
substantial way also in this simplified $\sigma-\omega$ model. This feature is
easily understood: Correlations, i.e.~the reduction of the NN wave functions at
short distances suppress the contributions of $\sigma$ and $\omega$ exchange to
a large extent. In order to understand the influence of the correlation effects
on the momentum dependence of the self-energy terms, one should realize that the
contributions to $G$ which are of second order in $V$ are attractive because of
the negative energy denominator in (\ref{eq:betheg}). The absolute values for
these energy denominators tend to decrease with increasing momentum of the
nucleon for which the self-energy is calculated. This means that the correlation
terms in $G$ tends to provide a more attractive contribution to the self-energy
components for nucleon momenta close to the Fermi momentum than for those with
momentum close to zero. This is just the effect which we see in
Fig.~\ref{fig:km35c} comparing the solid and the dashed lines. 

This momentum or better energy dependence of
contributions to the nucleon self-energy which are of second and higher order in
the NN interaction has also been observed  by Trasobares et al.~\cite{traso}, 
who studied the momentum and energy dependence of the nucleon self-energy in a
simple $\sigma - \omega$ model. It is very difficult to deduce this feature from
a Dirac analysis of the G matrix elements as it has been done in
\cite{horow,sehn,dejong,boer1} as such an analysis is based on an analysis of
the $G$ matrix in terms of a local interaction. The correlation effects which we
just discussed, however, yield non-local contributions to $G$.

\begin{figure}
\epsfysize=8.0cm
\begin{center}
\makebox[16.0cm][c]{\epsfbox{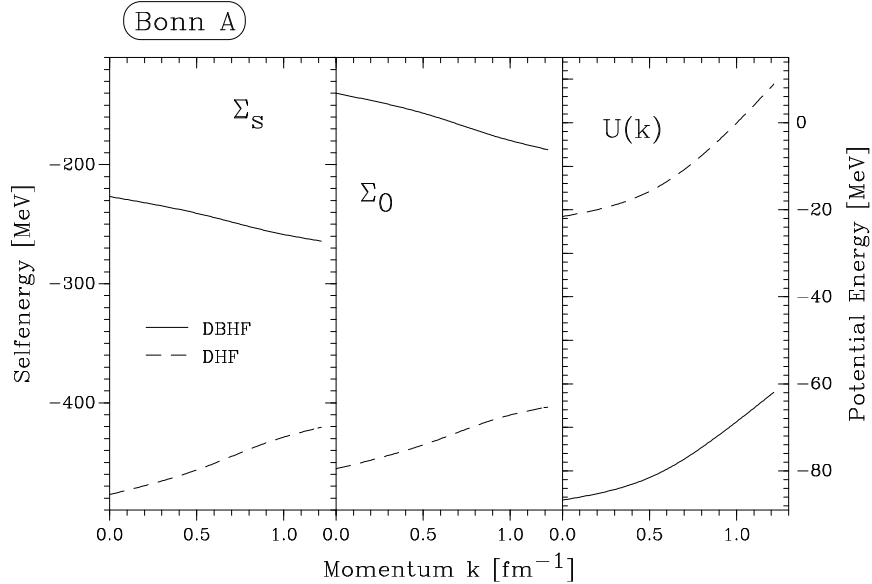}}
\end{center}
\caption{Self-energy and single-particle potential of nuclear matter with $k_F$
= 1.35 fm$^{-1}$ calculated for the Bonn $A$ potential. Further explanations see
Fig.~\protect{\ref{fig:km35c}}}
\label{fig:km35}
\end{figure}

In order to investigate the sensitivity of these results on the NN
interaction used, we show in Fig.~\ref{fig:km35} the various components of the
self-energy calculated at the same density as considered in Fig.~\ref{fig:km35c}
but using the version $A$ of the Bonn potential rather than $C$. The same
features are obtained for the NN interaction Bonn $A$ as we have discussed
before for Bonn $C$. The difference between these two potentials can be seen
best by comparing the results for single-particle potential $U(k)$ (see
eq.~(\ref{eq:uofk}) or eq.~(\ref{eq:ubhf})) which are given in the right wings
of figures \ref{fig:km35c} and \ref{fig:km35}. The DBHF results for this $U(k)$
range between -80 MeV and -60 MeV for nucleon momenta below the Fermi momentum
$k_F$ for both interactions considered. The contribution of the Born or HF term
is quite different in these two cases. While the Bonn $A$ yields values
between -20 MeV and +10 MeV, the HF results obtained for Bonn $C$ are between
+10 MeV and +40 MeV. This difference can be traced back to the different
strength of the tensor component in the NN interaction. The Bonn $C$ potential
has a stronger tensor component than Bonn $A$ (see discussion of the d-state
probabilities in the deuteron above). Therefore Bonn $C$ yields a larger 
attractive contribution originating from the iterated tensor force than Bonn
$A$. As both interactions were adjusted to fit NN scattering phase shifts, the
Born approximation to the scattering matrix, the bare potential $V$, must be
less attractive for Bonn $C$ than for $A$.

\begin{figure}
\epsfysize=9.0cm
\begin{center}
\makebox[16.0cm][c]{\epsfbox{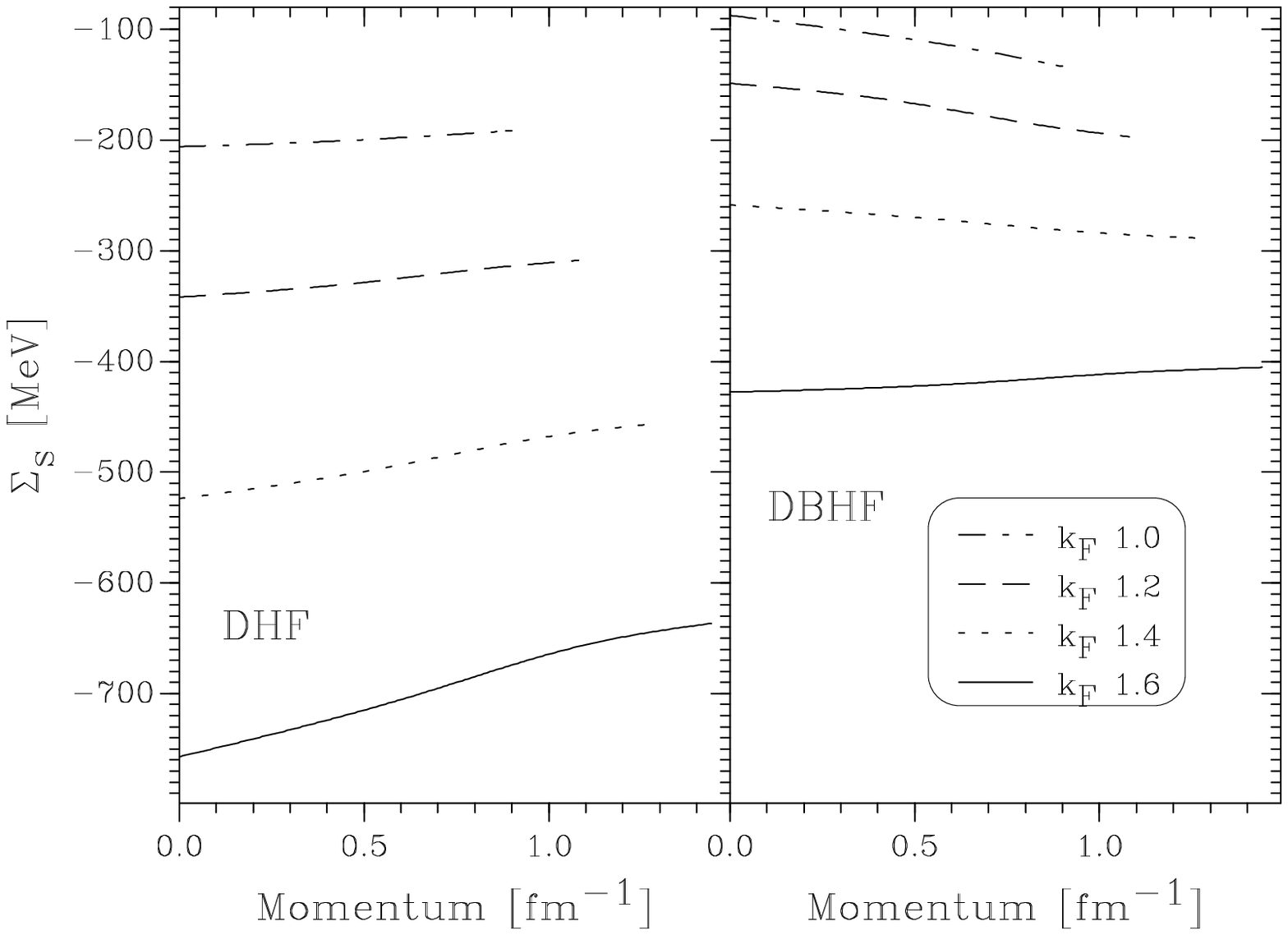}}
\end{center}
\caption{Momentum dependence of the scalar part, $\Sigma_s$ of the nucleon 
self-energy calculated for nuclear matter at various densities (as indicated by
the value of the Fermi momentum $k_F$) using Bonn $A$ interaction. The left part
of this figure contains the Born contribution, the right part the complete DBHF
result.}
\label{fig:sskma}
\end{figure}

The discussion above demonstrates that the  momentum dependence of the nucleon 
self-energy calculated in the DBHF approximation is a result of the Fock
contribution and effects of the correlation terms beyond the Born approximation
for the effective NN interaction, which tend to compensate each other to some
extent. This can also be seen from Fig.~\ref{fig:sskma}, which shows the
momentum dependence of the scalar component $\Sigma_s$ of the self-energy
calculated for nuclear matter of different densities, characterized by the
corresponding Fermi momentum. The left part of this figure contains the
self-energy contributions calculated in the Born approximation ($G=V$, denoted
by DHF) while the part on the right hand side shows the DBHF result obtained for
the complete $G$ matrix. For small densities the momentum dependence of
$\Sigma_s$ is dominated by the correlation term. For nuclear matter at higher
densities, however, the contributions to $G$ of second and higher order in $V$
are suppressed by the effects of the Pauli operator and the dispersive
corrections in the two-particle propagator of the Bethe-Goldstone
eq.~(\ref{eq:betheg}). Therefore in the examples visualized in
Fig.~\ref{fig:sskma} for the Fermi momentum $k_F$ = 1.6 fm$^{-1}$, the momentum
dependence of the Fock contribution overcompensates the momentum dependence
originating from the correlation effects and $\Sigma_s$ increases with momentum
also when calculated in the DBHF approach. Similar trends are also observed for
the momentum dependence of the vector component of the self-energy $\Sigma_0$.

\begin{figure}
\epsfysize=9.0cm
\begin{center}
\makebox[16.0cm][c]{\epsfbox{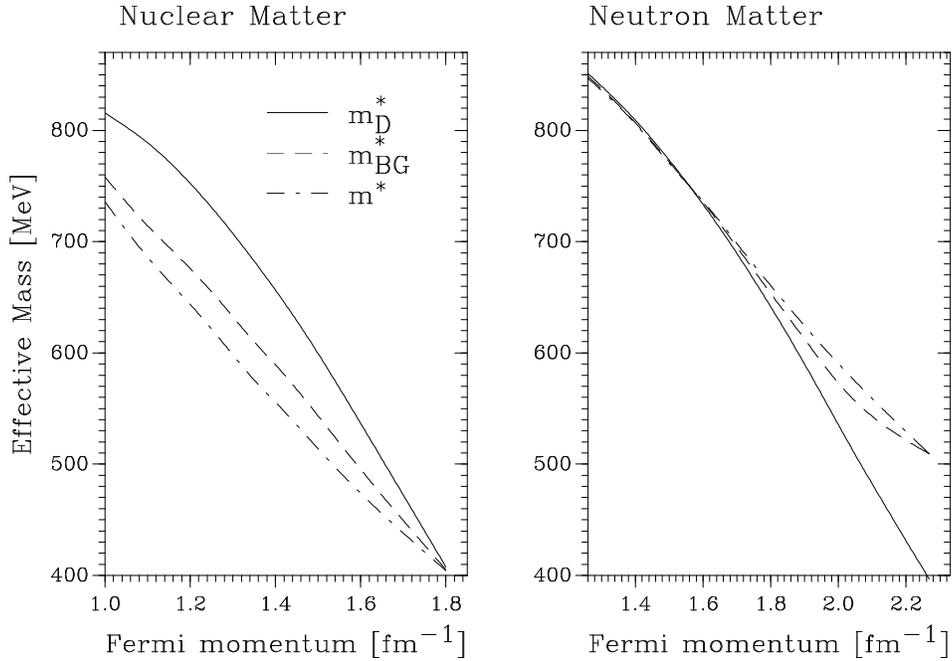}}
\end{center}
\caption{Various effective masses, as discussed in the text, calculated for 
nuclear matter (left part) and neutron matter using Bonn $A$ interaction. For 
the effective Dirac mass $m_D^*$, the value at $k=0.8 k_F$ is displayed. Results
are presented as a function of the Fermi momenta $k_F$.}
\label{fig:efma}
\end{figure}

In the next step we would like to study the effect of this improved analysis of
the nucleon self-energy on the self-consistent DBHF calculations. For that
purpose we present in Fig.~\ref{fig:efma} the value of the effective mass
$m_D^*$ characterizing the structure of the Dirac spinors (see 
eq.(\ref{eq:mdirac})) and the value of $m_{BG}^*$ (see eq.(\ref{eq:mbg}))
representing the single-particle spectrum in the Bethe-Goldstone equation as a
function of the Fermi momentum. One observes a non-negligible difference between
these two effective masses in particular at small densities. This reflects the 
importance of the momentum dependence of the self-energy which we just
discussed. These values are compared to the result for the
effective mass, $m^*$, derived in the simplified DH self-consistency scheme (see
eq.(\ref{eq:brok})). For densities around and below the empirical saturation
density of nuclear matter ($k_F$ = 1.35 fm$^{-1}$) the effective mass $m^*$
underestimates the values of both the Dirac mass $m_D^*$ and the
``Bethe-Goldstone mass'' $m_{BG}^*$ by a considerable amount. These three values
merge to essentially the same value only at densities above twice the empirical
saturation density.  

The situation is quite different for neutron matter, as can be seen from the
right part of Fig.~\ref{fig:efma}. Note that the range of Fermi momenta
displayed in this right part covers the same densities as displayed for nuclear
matter in the left part. All three effective masses coincide for neutron matter
at small densities. In this region, the momentum dependence of the Dirac
components of the self-energy is weaker than the momentum dependence in nuclear
matter at the same density. This can be traced back to a slight reduction of the
correlation effects in neutron matter as compared to nuclear matter. At large
densities the effective Dirac mass $m_D^*$ is smaller than $m_{BG}^*$ which is
close to $m^*$ derived in the DH approach to the DBHF self-consistency
requirement. It should be noted that the effective Dirac mass determined for
neutron matter is rather close to the value of $m_D^*$ derived for nuclear
matter at the same density, while deviations can be found for the effective 
masses describing the energy spectra. 

\begin{figure}
\epsfysize=9.0cm
\begin{center}
\makebox[16.0cm][c]{\epsfbox{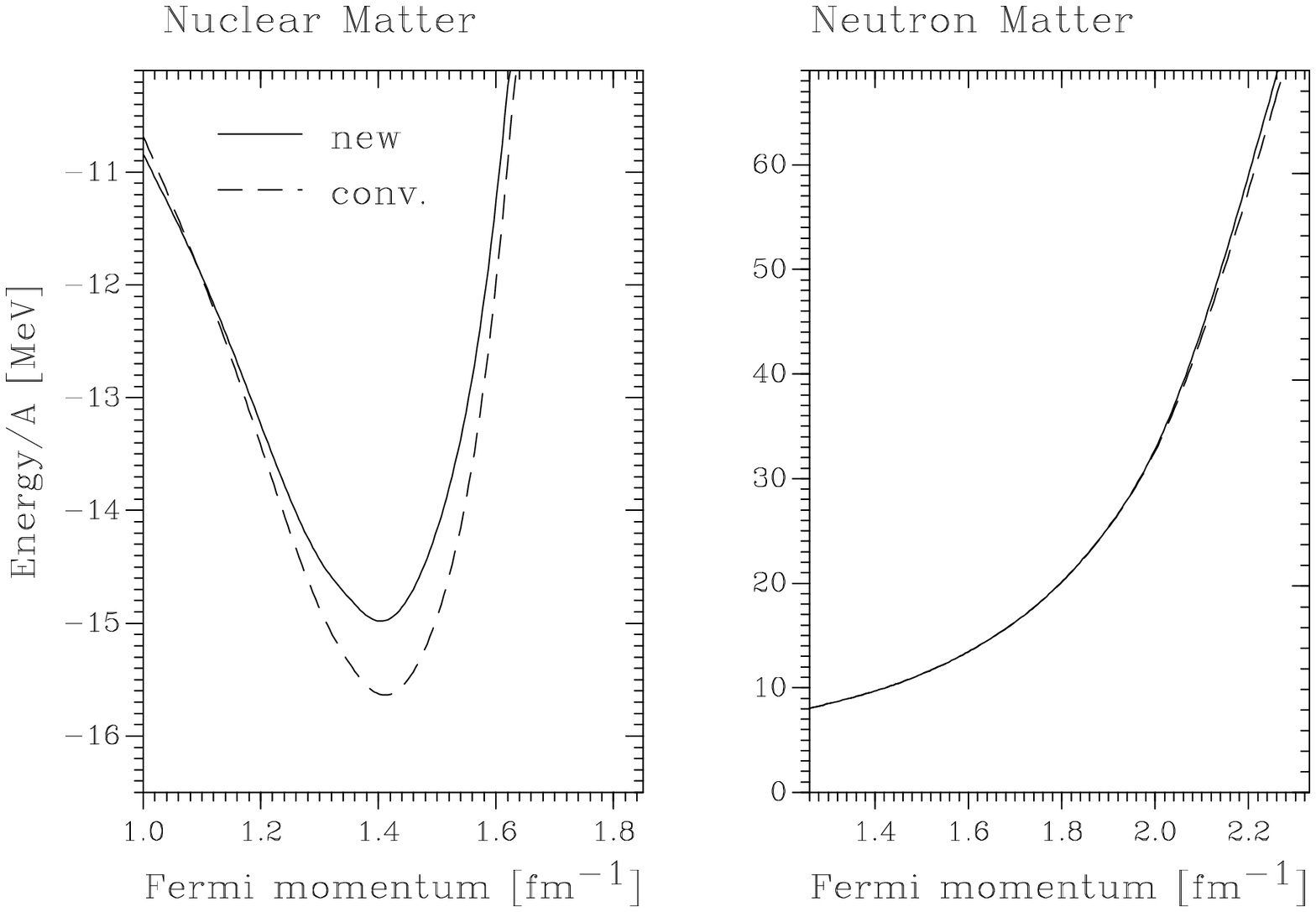}}
\end{center}
\caption{Energy of nuclear and neutron matter at various densities characterized
by the Fermi momentum $k_F$. Results obtained within the conventional DH
self-consistency scheme are represented by the dashed line, while those obtained
using the scheme which separates momentum dependence and Dirac structure are
shown using solid lines. Bonn $A$ potential has been used for the NN
interaction.}
\label{fig:enera}
\end{figure}

Although the effective masses derived within the DH self-consistency scheme
deviate considerably from those obtained within the improved scheme, the
calculated binding energies for nuclear matter and neutron matter are very close
to each other (see Fig.~\ref{fig:enera}). This is not very surprising. At small
densities the binding energies calculated within the DBHF and BHF approach,
i.e.~with and without accounting for any change of the Dirac spinors at all, are
close to each other (see e.g.~\cite{rupr,brock}), therefore one should not
expect any sensitivity of the calculated binding energies on the precise way of
calculating the effective Dirac mass at these densities. At large densities of
nuclear matter the DH scheme and its improvement yield results which are close
to each other, therefore the two methods should not lead to any substantial
differences in this area. Deviations are only visible at densities around the
saturation point, where the new scheme predicts a binding energy which is
slightly smaller than obtaiend within the old scheme. 

We expect that these differences will show up in
calculations of finite nuclei which are based on nuclear matter calculations at
such small densities using a local density
approximation\cite{fritz,toki1,toki2,boer2}. For such calculations one defines
an effective meson theory, typically considering the exchange of a scalar,
$\sigma$ and a vector meson $\omega$ only, adjusting the coupling constants at
each density in such a way that this effective meson exchange model used within
the Dirac-Hartree approach reproduces some key quantities of the DBHF
calculation of nuclear matter. For these key quantities to be fitted we will 
use here the value of $m_D^*$ and the total binding energy. 

\begin{figure}
\epsfysize=9.0cm
\begin{center}
\makebox[16.0cm][c]{\epsfbox{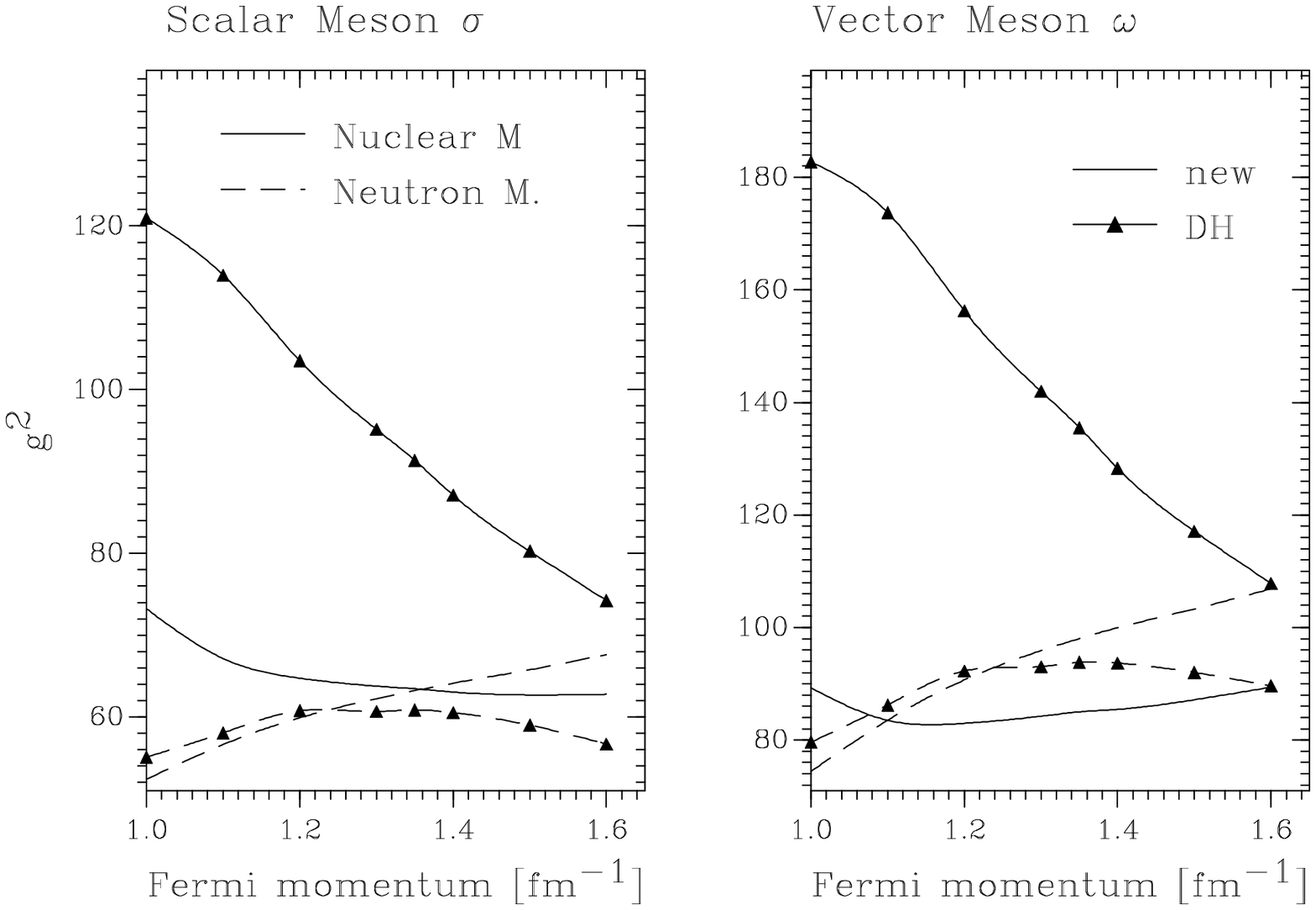}}
\end{center}
\caption{Effective Coupling constants for a scalar meson ($\sigma$, left part of
the figure) and vector meson ($\omega$, right part of the figure). The coupling
constants were adjusted at each density in such a way that a Dirac Hartree
calculation using these coupling constants fits the Dirac mass $m_D^*$ and the
calculated energy of the DBHF calculation. Solid lines refer to nuclear matter,
while the dashed lines are obtained for neutron matter. The lines marked with
triangles are derived from DBHF calculations using the old DH scheme. The other
lines are obtained from the calculations with effective Dirac masses derived
from eq.(\protect\ref{eq:mdsts}). The meson masses are 550 MeV and
783 MeV for the $\sigma$ and $\omega$, respectively.}
\label{fig:efcoa}
\end{figure}

It has been a problem of such investigations based on DBHF using the old DH
self-consistency scheme that the results were sensitive to the extrapolation of the
nuclear matter results to small densities, for which no stable nuclear matter
calculation can be performed. The reason for this sensitivity is exhibited in
Fig.~\ref{fig:efcoa}. The effective coupling constants for $\sigma$ and $\omega$
derived within this scheme are getting larger for nuclear matter with decreasing 
density. If, however, the self-consistency scheme is employed, which separates
the Dirac structure and the momentum dependence of the self-energy, one obtains
effective coupling constants, which are weaker and depend much less on the
nuclear density. Also the isospin dependence is weaker as can be seen from the
comparison between nuclear matter and neutron matter. Therefore we expect more
reliable results for finite nuclei using the effective coupling constants
derived from these new DBHF calculations than were obtained within the old
scheme. This should particularly be true for asymmetric
systems\cite{engvi,toki2}.

\section{Conclusions}
The single-particle potential of nucleons in nuclear matter, calculated in the 
framework of the Dirac Brueckner Hartree Fock (DBHF) approach, is non-local or
momentum dependent. This momentum dependence is conveniently expressed in terms
of effective masses of the nucleon. Possible sources for this non-locality  are
the Fock exchange terms, the nucleon-nucleon correlations accounted for by the
Brueckner G matrix and the relativistic structure of the self-energy, i.e. its
decomposition into  Lorentz scalar and vector terms. It is the aim of the work
presented here to explore the importance of these different sources and their
mutual influence. In other words we want to explore the contribution of these
sources to the effective mass of the nucleon. 

For that  purpose a new method has been presented to enable the decomposition
of the self-energy into its various Dirac components. It is based on the
dependence of the single-particle energy on the small component in the Dirac
spinors used to calculate the matrix elements of the underlying NN interaction.
This technique has successfully been tested for the case of the Dirac Hartree
Fock (DHF) approach, for which the Dirac components can also be calculated 
directly. The comparison between DHF and DBHF results demonstrates the
importance of the NN correlations taken into account in terms of the Brueckner
G matrix. These correlations tend to reduce the absolute values of the self
energies by almost a factor of two. The reduction is slightly larger for the
vector component than for the scalar one, which leads to a bound system of
nuclear matter, while the DHF approach using the bare Bonn potentials leads to
unbound nuclear matter.

The correlation and non-locality effects contained in the Brueckner G matrix
are also very important to understand the momentum dependence of scalar and
vector components of the self-energy. At densities around and below the
saturation  density of nuclear matter the contributions to  G  beyond the Born
approximation yield a momentum dependence of these components which is opposite
to the one derived in the DHF approximation. At higher densities these
correlation effects are suppressed and the momentum dependence due to the Fock
exchange terms is getting more  important. The importance of these nonlocal
contributions to G cast some doubt on the validity of so-called projection
scheme which derives the Dirac structure of the self-energy from an expansion
of the two-body interaction, assuming a local parametrization.

The effective mass of the nucleon in pure neutron matter is dominated by the
relativistic effects, the effective mass describing the momentum dependence of
the single-particle potential is essentially identical to the Dirac mass
$m_D^*$.  The same is true for nuclear matter at high densities. At smaller
densities, however, the effective mass parametrizing the single-particle
potential is significantly  smaller than $m_D^*$. The Dirac masses for nuclear
matter and neutron matter at the same density are rather close to each other.
This reflects the fact that the contribution to the NN interaction obtained
from the exchange of a scalar isovector meson, like the $\delta$ meson, is much
weaker than the contribution of the scalar isoscalar ($\sigma$) meson.

The improved analysis of the Dirac structure of the nucleon self-energy has
only little effect on the calculated binding energy of nuclear matter. The
deviations between the simple Dirac Hartree analysis and the present scheme are
largest at densities below the saturation density of nuclear matter. Therefore
we expect larger differences in the calculation of finite nuclei. 

This work has been supported by the Sonderforschungsbereich SFB 382 of the
``Deutsche Forschungsgemeinschaft'' and by the ``Humboldt-Stiftung''.

\vfil\eject

\begin{thebibliography}{99}
\bibitem{wal1} J.D.\ Walecka, Ann.\ Phys.\ (N.Y.) {\bf 83}, 491 (1974).
\bibitem{serot} B.D.\ Serot and J.D.\ Walecka, Adv.\ Nucl.\ Phys.\ {\bf 16}, 1
(1986).
\bibitem{anast} M.R.\ Anastasio, L.S.\ Celenza, W.S.\ Pong, and C.M.\ Shakin,
Phys.\ Rep.\ {\bf 100}, 327 (1983).
\bibitem{rupr} R.\ Machleidt, Adv.\ Nucl.\ Phys.\ {\bf 19}, 189 (1989).
\bibitem{brock} R.\ Brockmann and R.\ Machleidt, Phys.\ Lett.\ {\bf B149}, 283
(1984).
\bibitem{bm2} R.\ Brockmann and R.\ Machleidt, Phys.\ Rev. {\bf C42}, 1965
(1990).
\bibitem{malf1} B.\ Ter Haar and R.\ Malfliet, Phys.\ Rep.\ {\bf 149}, 207
(1987).
\bibitem{weigel} H.\ Huber, F.\ Weber, and M.K.\ Weigel, Phys.\ Lett.\ {\bf
B317}, 485 (1993).
\bibitem{bwi} R.B.\ Wiringa, V.\ Fiks and A.\ Fabrocini, Phys.\ Rev.\ {\bf C38},
1010 (1988).
\bibitem{engvi} L.\ Engvik, M.\ Hjorth-Jensen, E.\ Osnes, G.\ Bao and
E.\ \O stgaard, Phys.\ Rev.\ Lett.\ {\bf 73}, 2650 (1994).
\bibitem{lee} C.H.\ Lee, T.T.S.\ Kuo, G.Q.\ Li and G.E.\ Brown, Preprint,
nucl-th/9705012.
\bibitem{jami} M.\ Jaminon and C.\ Mahaux, Phys.\ Rev.\ {\bf C41}, 697 (1990).
\bibitem{horow} C.J.\ Horowitz and B.D.\ Serot, Nucl.\ Phys.\ {\bf A464}, 613
(1987).
\bibitem{sehn} L.\ Sehn, C.\ Fuchs and A.\ Faessler, Phys.\ Rev.\ {\bf C56}, 216
(1997).
\bibitem{dejong} F.\ de Jong and H.\ Lenske, Preprint, nucl-th/9707017.
\bibitem{fritz} R.\ Fritz and H.\ M\"uther, Phys.\ Rev.\ {\bf C49}, 633 (1994).
\bibitem{toki1} R.\ Brockmann and H.\ Toki, Phys.\ Rev.\ Lett.\ {\bf 68}, 3408
(1992).
\bibitem{toki2} H.\ Shen, Y.\ Sugahara, and H.\ Toki, Phys.\ Rev.\ {\bf C55},
1211 (1997).
\bibitem{boer1} H.F.\ Boersma and R.\ Malfliet, Phys.\ Rev.\ {\bf C49}, 233
(1994), Erratum {\bf C50}, 1253 (1994).
\bibitem{boer2} H.F.\ Boersma and R.\ Malfliet, Phys.\ Rev.\ {\bf C49}, 1495
(1994).
\bibitem{ulr1} S.\ Ulrych and H.\ M\"uther, Phys.\ Rev.\ {\bf C} in press,
nucl-th/9706030.
\bibitem{elsen} H.\ Elsenhans, H.\ M\"uther and R.\ Machleidt, Nucl.\ Phys.\
{\bf A515}, 715 (1990).
\bibitem{traso} A.\ Trasobares, A.\ Polls, A.\ Ramos, and H.\ M\"uther ,
Preprint nucl-th/9703020.
\end{thebibliography}
\end{document}